\documentclass[12pt]{JHEP3}

\usepackage{epsfig}

\def\spose#1{\hbox to 0pt{#1\hss}}
\def\ltapprox{\mathrel{\spose{\lower 3pt\hbox{$\mathchar"218$}}
 \raise 2.0pt\hbox{$\mathchar"13C$}}}
\def\gtapprox{\mathrel{\spose{\lower 3pt\hbox{$\mathchar"218$}}
 \raise 2.0pt\hbox{$\mathchar"13E$}}}

\title{
$\theta$ dependence of SU($N$) gauge theories
}

\author{Luigi Del Debbio \\
	Dipartimento di Fisica dell'Universit\`a 
	di Pisa and I.N.F.N. \\
        Via Buonarroti 2, I-56127 Pisa, Italy \\ 
	E-mail: \email{ldd@df.unipi.it} 
} 

\author{Haralambos Panagopoulos \\ 
        Department of Physics, University of Cyprus\\
        Lefkosia, CY-1678, Cyprus\\ 
	E-mail: \email{haris@ucy.ac.cy} 
}

\author{Ettore Vicari \\ 
	Dipartimento di Fisica dell'Universit\`a 
	di Pisa and I.N.F.N. \\
        Via Buonarroti 2, I-56127 Pisa, Italy \\ 
	E-mail: \email{vicari@df.unipi.it} 
}

\abstract{ We study the $\theta$ dependence of four-dimensional
SU($N$) gauge theories, for $N\geq 3$ and in the large-$N$ limit.  We
use numerical simulations of the Wilson lattice formulation of gauge
theories to compute the first few terms of the expansion of the
ground-state energy $F(\theta)$ around $\theta=0$, $F(\theta)-F(0) =
A_2 \,\theta^2 ( 1 + b_2 \theta^2 + ...)$.  
Our results support
Witten's conjecture: $F(\theta)-F(0) = {\cal A}\,\theta^2 +
O(1/N)$ for sufficiently small values of $\theta$, $\theta < \pi$.
Indeed we verify that the topological susceptibility has a nonzero
large-$N$ limit $\chi_\infty=2 {\cal A}$ with corrections of
$O(1/N^2)$, in substantial agreement with the Witten-Veneziano formula
which relates $\chi_\infty$ to the $\eta^\prime$ mass. Furthermore,
higher order terms in $\theta$ are suppressed; in particular, the
$O(\theta^4)$ term $b_2$ (related to the $\eta^\prime - \eta^\prime$
elastic scattering amplitude) turns out to be quite small:
$b_2=-0.023(7)$ for $N=3$, and its absolute value decreases with
increasing $N$, consistently with the expectation $b_2=O(1/N^2)$.  }

\begin{document}

\section{Introduction}

It is rather well established that four-dimensional SU($N$) gauge theories
must have a nontrivial dependence on the angle $\theta$ 
that appears in the Euclidean Lagrangian as
\begin{equation}
{\cal L}_\theta  = {1\over 4} F_{\mu\nu}^a(x)F_{\mu\nu}^a(x)
- i \theta q(x)
\label{lagrangian}
\end{equation}
where $q(x)$ is the topological charge density
\begin{equation}
q(x) = {g^2\over 64\pi^2} \epsilon_{\mu\nu\rho\sigma}
F_{\mu\nu}^a(x)F_{\rho\sigma}^a(x).
\end{equation}
Indeed, the most plausible explanation of how the solution of the
so-called U(1)$_A$ problem can be compatible with the $1/N$ expansion
(performed keeping $g^2N$ fixed \cite{Hooft-74}) suggests that in the
pure gauge theory the ground-state energy depends on $\theta$ to
leading order in $1/N$ \cite{Witten-79}.  Numerical nonperturbative
studies, based on Monte Carlo simulations of the lattice formulation of
SU($N$) gauge theories (see
e.g. Ref.~\cite{Teper-00,GarciaPerez:2001hq} for recent reviews), have
given nonzero estimates for the topological susceptibility $\chi$ at
$\theta=0$, which is the zero-momentum two-point correlation function
of $q(x)$, and therefore the second derivative of the ground state
energy with respect to $\theta$ computed at $\theta=0$.

The $\theta$ dependence of the theory is described
by the partition function 
\begin{equation}
Z(\theta) = \int [dA] \exp \left(  - \int d^4 x {\cal L}_\theta \right)
\equiv \exp[ - V F(\theta) ]
\end{equation}
where $F(\theta)$ is the free energy (ground state energy) per 
unit space-time volume.
Some information on the behavior of $F(\theta)$ comes
from large-$N$ arguments.
Witten argued \cite{Witten-80} that in the large-$N$ limit
$F(\theta)$ is a multibranched function of the type
\begin{equation}
F(\theta) = N^2 {\rm min}_k\, H\left( {\theta+2\pi k\over N}\right)
\label{conj1}
\end{equation}
which is periodic in $\theta$, but not smooth
since at some value of $\theta$ there is a jump between two different
branches. 
More recently, the conjecture was refined \cite{Witten-98}
leading to a rather simple expression for $\Delta F(\theta)\equiv
F(\theta)-F(0)$ in the large-$N$ limit, that is
\begin{equation}
\Delta F(\theta) = {\cal A} \, {\rm min}_k \, (\theta+2\pi k)^2 + O\left(
1/N\right).
\label{conj2}
\end{equation} 
In particular, for sufficiently small values of $\theta$,
i.e. $\theta<\pi$, 
\begin{equation}
\Delta F(\theta) = {\cal A} \, \theta^2  + O\left( 1/N\right).
\label{conj2b}
\end{equation} 
Thus possible $O(\theta^4)$ terms are expected to be depressed
by powers of $1/N$.
This conjecture has been supported using arguments 
based on a duality of large-$N$ gauge theories
with string theory \cite{Witten-98}.
It has also been discussed in a field-theoretical framework
in Ref.~\cite{Gabadadze-99}.

In this work we study the $\theta$ dependence
of SU($N$) gauge theories, using numerical simulations
of their Wilson lattice formulation.
Numerical Monte Carlo studies of the $\theta$-dependence are made 
very difficult by the complex nature of the $\theta$ term.
In fact the lattice action corresponding to the Lagrangian
(\ref{lagrangian}) cannot be directly simulated
for $\theta\ne 0$,
not to mention the difficulty to define a topological charge
density on the lattice. 
Here we restrict ourselves to the region of
small $\theta$ values, 
where one may expand $F(\theta)$ around $\theta=0$
\begin{equation}
\Delta F(\theta) = \sum_{i=1} A_{2i} \theta^{2i}.
\label{expf}
\end{equation}
The coefficient $A_2$ gives the topological susceptibility, i.e. 
\begin{equation}
A_{2} = {1\over 2} \chi, \qquad \chi\equiv\chi_2 
= {1\over V} \langle Q^2 \rangle_{\theta=0},
\end{equation}
where $Q$ is the topological charge $Q=\int d^4 x\, q(x)$.
Higher order coefficients of the expansion (\ref{expf})
can be related to higher moments of the probability distribution
$p(Q)$ of the topological charge.
For instance,
\begin{eqnarray}
&&A_4 = - {1\over 24} \chi_4,\nonumber \\
&& \chi_4 = {1\over V} \left[ 
\langle Q^4 \rangle_{\theta=0} - 3 \left( 
\langle Q^2 \rangle_{\theta=0} \right)^2 \right]. \label{chi4}
\end{eqnarray}
A convenient parametrization of the free energy is
achieved by introducing the dimensionless function
\begin{equation}
f(\theta) = \sigma^{-2}\,\Delta F(\theta),
\end{equation}
where $\sigma$ is the string tension at $\theta=0$.
The coefficients of the $\theta$-expansion of $f(\theta)$
are dimensionless scaling quantities. Let us write $f(\theta)$ in the
form
\begin{equation}
f(\theta) = {1\over 2} C \,\theta^2 s(\theta)
\label{sdef}
\end{equation} 
where 
\begin{eqnarray}
&&C={2 A_2\over \sigma^2} = {\chi\over \sigma^2}, \label{Cexp}\\
&& s(\theta) = 1 + b_{2} \theta^2 + b_4 \theta^4 + ... ,\label{sexp}\\
&&b_2 = {A_4\over A_2} = - {\chi_4\over 12\chi},\label{b2exp}
\end{eqnarray}
The function $s(\theta)$ parametrizes the corrections 
to a simple Gaussian behavior.
Witten's conjecture implies that
$C$ has  a finite nonzero large-$N$ limit $C_{\infty}$
with $O(1/N^2)$ corrections, 
and $b_{2i}=O(1/N^2)$.
In particular one expects that $b_2$ is small, and that it should
rapidly decrease with increasing $N$.
We recall that the nonzero value of $\chi_{\infty}$ is essential
to provide an explanation to the U(1)$_A$ problem, and can be related 
to the $\eta'$ mass \cite{Witten-79,Veneziano-79} through the relation
\begin{equation}
\chi_\infty = 2 {\cal A} = {f_\pi^2 m_{\eta'}^2\over 4 N_f} + O(1/N).
\label{wittenformula}
\end{equation}
The quantity $b_2$ also lends itself to a physical interpretation,
being related to the $\eta^\prime - \eta^\prime$ elastic scattering
amplitude \cite{Veneziano-79}.

We present results for four-dimensional
SU($N$) gauge theories with $N=3,4,6$. In particular, we determine 
the ratio $C=\chi/\sigma^2$ and the 
coefficient $b_2$ of the $O(\theta^2)$ term in the 
function $s(\theta)$. 
Once verified that the results for $N=3,4,6$ are consistent with
the expected behavior of the type $A+B/N^2$,
the large-$N$ limit is determined by a fit.
This allows us to obtain a rather accurate estimate
of $C_\infty$. 
The estimated values of $b_2$ turn out to be very small, of the order
of a few per cent, and appear to decrease rapidly with $N$,
supporting Witten's conjecture.
Our results provide a first quantitative
study of the function $s(\theta)$, cf.  Eq.~(\ref{sdef}),
related to the corrections to the $\theta^2$ term in the
free energy $F(\theta)$, i.e. to a simple Gaussian behavior.

The paper is organized as follows. In Sec.~\ref{sec2}
we describe the Monte Carlo simulations that we have performed.
In Sec.~\ref{sec3} we present and discuss the results.

\section{Numerical simulations}
\label{sec2}

\TABLE[ht]{
\caption{
Results of our Monte Carlo simulations for $N=3$.
We tabulate the coupling $\beta$,
the corresponding value of the string tension 
(taken from Ref.~\protect\cite{LT-01} for $N=3$),
the lattice size $L$, 
the number of sweeps of the simulation $N_{\rm sw}$,
the autocorrelation time $\tau_Q$ of the topological charge,
the topological susceptibility  $\chi$, the scaling ratio
$C=\chi/\sigma^2$ and the coefficient $b_2$.
The error on the ratio $C$ is computed by considering
the errors on $\chi$ and $\sigma$ as independent.
}
\label{tableres3}
\begin{tabular}{llccclll}
\hline\hline
\multicolumn{1}{c}{$\beta$}&
\multicolumn{1}{c}{$\sqrt{\sigma}$}&
\multicolumn{1}{c}{$L$}&
\multicolumn{1}{c}{$N_{\rm sw}/10^6$}&
\multicolumn{1}{c}{$\tau_Q$}&
\multicolumn{1}{c}{$10^4\,\chi$}&
\multicolumn{1}{c}{$C$}&
\multicolumn{1}{c}{$b_2$}\\
\hline \hline
$5.9$ & 0.2605(14) & $16$ & 8.7 & 29(1) & 1.549(12) & 0.0336(8) & $-$0.028(12) \\ 
             &            & $12$ & 5.0 &       & 1.544(7)  & 0.0335(7) & $-$0.021(5)  \\

   $6.0$ & 0.2197(12) &$16$ &  6.4 & 77(5) & 0.741(5)  & 0.0318(7) & $-$0.023(7)\\
               &            &$12$ &  4.0 & 70(1) & 0.728(5)  & 0.0312(7) & $-$0.032(3)\\

   $6.1$ & 0.1876(12) &$16$ &  1.6 &173(4) & 0.382(6)  & 0.0308(9) & $-$0.032(9) \\ 

   $6.2$ & 0.1601(10) &$20$ &  6.9 &540(30)& 0.1989(23)& 0.0303(8) & $-$0.016(8)  \\
\hline\hline
\end{tabular}
}

\TABLE[ht]{
\caption{
Results of our Monte Carlo simulations for $N=4$.
We tabulate $\gamma\equiv \beta/(2 N^2)$,
the corresponding value of the string tension 
(taken from Ref.~\protect\cite{DPRV-02}),
the lattice size $L$ (in most cases we combine results obtained
for $L^4$ and $L^3\times 2L$ lattices), 
the number of sweeps of the simulation $N_{\rm sw}$,
the autocorrelation time $\tau_Q$ of the topological charge,
the topological susceptibility  $\chi$, the scaling ratio
$C=\chi/\sigma^2$ and the coefficient $b_2$.
The error on the ratio $C$ is computed by considering
the errors on $\chi$ and $\sigma$ as independent.
}
\label{tableres4}
\begin{tabular}{llccclll}
\hline\hline
\multicolumn{1}{c}{$\gamma$}&
\multicolumn{1}{c}{$\sqrt{\sigma}$}&
\multicolumn{1}{c}{$L$}&
\multicolumn{1}{c}{$N_{\rm sw}/10^6$}&
\multicolumn{1}{c}{$\tau_Q$}&
\multicolumn{1}{c}{$10^4\,\chi$}&
\multicolumn{1}{c}{$C$}&
\multicolumn{1}{c}{$b_2$}\\
\hline \hline
$0.335$& 0.2959(14)&$12$ &  2.0 &  & 2.274(22) & 0.0297(6) & $-$0.04(3) \\

  $0.337$& 0.2699(23)&$16$ & 2.3 & 85(4)  & 1.675(16) & 0.0316(11)& $-$0.01(5) \\

  $0.338$& 0.2642(7) &$12$ & 7.9 & 105(4) & 1.416(9)  & 0.0291(4) & $-$0.012(6) \\ 

  $0.341$& 0.2368(6) &$12$ & 6.4 & 207(6) & 0.896(10) & 0.0285(4) & $-$0.013(6) \\ 

  $0.344$& 0.2160(8) &$16$ & 6.8 & 410(20)& 0.608(7)  & 0.0279(5) & $-$0.016(12) \\
                &           &$12$ & 6.1 & 420(20)& 0.595(18) & 0.0273(9) & $-$0.019(7) \\

  $0.347$& 0.1981(5) &$16$ & 10.2 &805(50)&0.425(10) & 0.0276(7) & $-$0.011(8) \\ 
\hline\hline
\end{tabular}
}

\TABLE[ht]{
\caption{
Results of our Monte Carlo simulations for $N=6$.
All tabulated quantities are as in Table~\ref{tableres4}.
}
\label{tableres6}
\begin{tabular}{llcccll}
\hline\hline
\multicolumn{1}{c}{$\gamma$}&
\multicolumn{1}{c}{$\sqrt{\sigma}$}&
\multicolumn{1}{c}{$L$}&
\multicolumn{1}{c}{$N_{\rm sw}/10^6$}&
\multicolumn{1}{c}{$\tau_Q$}&
\multicolumn{1}{c}{$10^4\,\chi$}&
\multicolumn{1}{c}{$C$}\\
\hline \hline
$0.342$ &0.3239(8) &$12$ & 0.5 & 250(25) & 2.55(7)   & 0.0232(7) \\ 

$0.344$ &0.2973(5) &$12$ & 0.7 & 466(50) & 1.79(6)   & 0.0228(8)  \\ 

$0.348$ & 0.2535(6)&$16 $ &  1.2 & 2675(500)  & 0.91(7)   & 0.0220(17)  \\ 

  &&                         $12$ & 4.1 & 3080(120)
   & 1.01(5)   & 0.0244(12) \\ 

  &&                         $10$ & 0.6 &   & 0.99(10)  & 0.0240(24) \\ 

  $0.350$ & 0.2380(6)&$12$ & 0.5 & $\gtrsim 3500$& 0.76(9)   & 0.0237(28)  \\ 
\hline\hline
\end{tabular}
}

We consider the Wilson formulation of lattice gauge theories:
\begin{equation}
S = - N\beta \sum_{x,\mu>\nu} {\rm Tr} \left[
U_\mu(x) U_\nu(x+\mu) U_\mu^\dagger(x+\nu) U_\nu^\dagger(x) 
+ {\rm h.c.}\right],
\label{wilsonac}
\end{equation}
where $U_\mu(x)\in$ SU($N$) are link variables.
In our simulations we employed the Cabibbo-Marinari algorithm
\cite{CM-82} to upgrade SU($N$) matrices by updating their SU(2)
subgroups (we selected $N(N-1)/2$ subgroups and 
each matrix upgrading consists of $N(N-1)/2$ SU(2) updatings). 
This was done by alternating
microcanonical over-relaxation and heat-bath steps, typically in a 4:1
ratio. Tables~\ref{tableres3}, \ref{tableres4} and \ref{tableres6}
contain some information on our MC runs respectively for $N=3,4,6$:
The coupling values\footnote{
In order to compare Monte Carlo results for various values of $N$, it
is useful to introduce the rescaled coupling
$\gamma \equiv {1/ (g_0^2 N)} = {\beta / (2 N^2)}$, which is kept fixed
when taking the large-$N$ limit of the lattice theory.}
$\beta$ or $\gamma\equiv \beta/(2 N^2)$,
the corresponding value of the string tension (taken from Refs.~\cite{DPRV-02,LT-01}),
lattice sizes and the number of sweeps. 
The values of $\beta,\gamma$ were chosen to lie in the weak-coupling region,
i.e., beyond the first order phase transition in the case $N=6$,
and beyond the crossover region characterized by a peak of the
specific heat for $N=3,4$;
see Ref.~\cite{DPRV-02} for a more detailed discussion of this point.
Computing quantities related to topology using
lattice simulation techniques is not a simple task.
In a lattice theory the fields are defined on a discretized
set, therefore the topological properties are strictly trivial.
One relies on the fact that the physical topological properties
are recovered in the continuum limit.
Various techniques have been proposed and employed to
associate a topological charge to a lattice configuration,
see e.g. the review \cite{Teper-00}.
Here we employ a cooling technique.
The topological charge has been measured on cooled configurations,
using the standard twisted double plaquette operator.
As is well known, the sum over the whole lattice of the twisted double
plaquette, measured on cooled 
configurations, takes on values $Q_t \simeq k \alpha$, where $k$ is an integer
and $\alpha \lesssim 1$.
We determine the typical value of 
$\alpha$ by minimizing the average deviation of
the twisted double plaquette from integer multiples of $\alpha$, and then
assign to $Q$ the integer closest to  $Q_t/\alpha$. 
This method eliminates the need for
expensive, protracted cooling; 
usually 
fewer than 20 steps suffice (when using $N(N-1)/2$ subgroups) in order
to observe a substantial convergence of the results, which become
indistinguishable from results obtained using either improved
operators for $Q$ or more cooling steps.
We measured the topological charge $Q$ typically every 100 sweeps.

As already observed in Ref.~\cite{DPRV-02},
a severe form of critical slowing down affects the measurement of $Q$,
posing a serious limitation for numerical studies of the
topological properties in the continuum limit, especially 
at large values of $N$.
In Tables~\ref{tableres3}, \ref{tableres4}  and \ref{tableres6}
we report the estimates of
the autocorrelation time $\tau_Q$ obtained  
by a blocking analysis of the data, where
the quoted errors on the estimates of $\tau_Q$ are just indicative,
and attempt to quantify the uncertainty in extracting
$\tau_Q$ from the blocking procedure outlined in  Ref.~\cite{DPRV-02}.
These results are suggestive of an interesting phenomenon: 
As shown in Fig.~\ref{tau}, the data turn out to be
well reproduced by an exponential behavior of the type
\begin{equation}
\ln \tau_Q \approx N \left( e_N \xi_\sigma + c_N \right)
\label{tauexp}
\end{equation}
where $\xi_\sigma$ is the length scale associated with
the string tension, i.e. $\xi_\sigma\equiv \sigma^{-1/2}$,
and $e_N$, $c_N$ are constants.
A fit to the data for $N=3,4,6$, 
using the function (\ref{tauexp}), gives respectively
$e_3 = 0.40(1) $ and $c_3 = -0.42(4)$,
$e_4 = 0.42(1) $ and $c_4 = -0.43(4)$, 
$e_6 = 0.48(3) $ and $c_6 = -0.55(9)$,
showing that
$\ln \tau_Q$ is approximately proportional to $N$ (at fixed $\xi_\sigma$).
A qualitative explanation of this severe form of critical slowing down 
may be that topological modes give rise to sizeable free-energy barriers
separating different regions of the configuration space.  
As a consequence, the
evolution in the configuration space may present a long-time
relaxation due to transitions between different topological charge
sectors.   
We do not have a priori arguments in favor
of the exponential behavior (\ref{tauexp}), which is just a guess 
arising naturally after observing the plot in Fig.~\ref{tau}.
Note also that this dramatic effect is not observed 
in plaquette-plaquette or Polyakov line correlations,
suggesting an approximate
decoupling between topological modes and nontopological ones,
such as those determining the confining properties.  

\FIGURE[ht]{
\epsfig{file=tau.eps, width=12truecm} 
\caption{
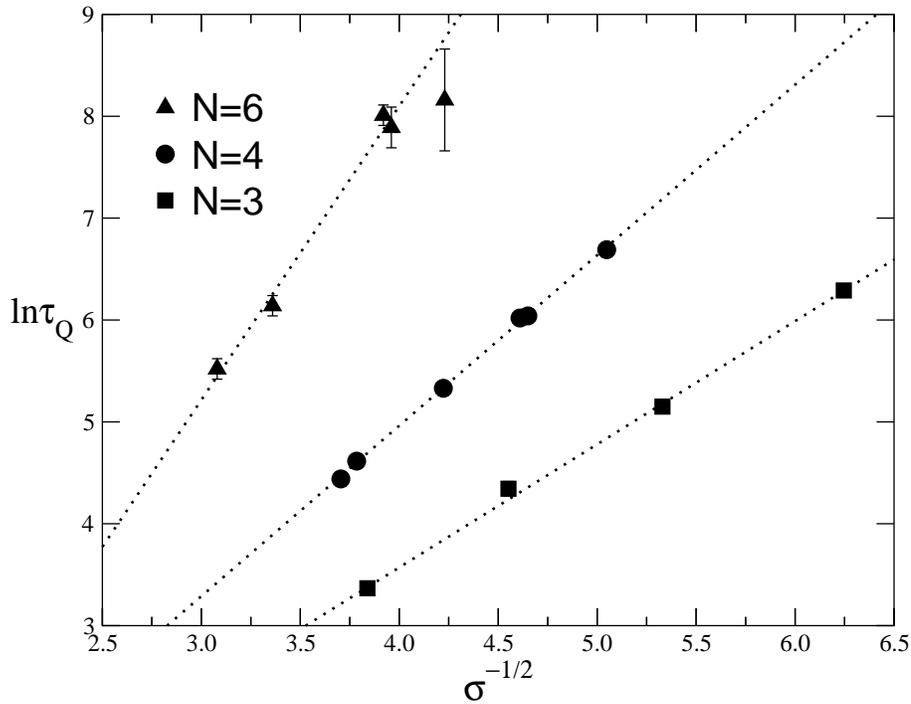
The autocorrelation time $\tau_Q$ versus $\xi_\sigma\equiv
\sigma^{-1/2}$. The dotted lines show the linear fits.
}
\label{tau}
}

\FIGURE[ht]{
\epsfig{file=chi2.eps, width=12truecm} 
\caption{
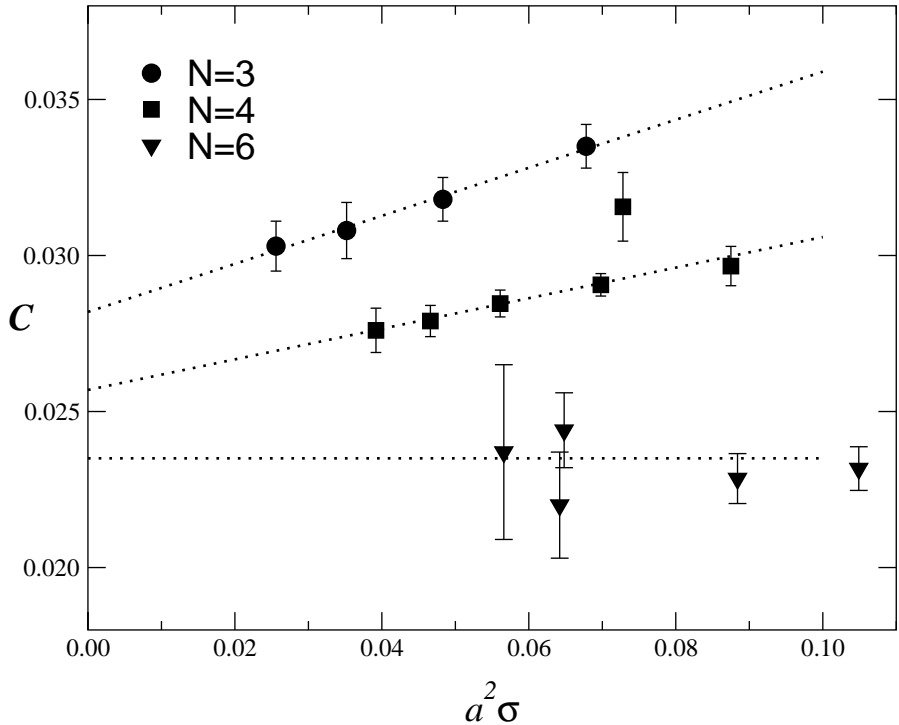
The scaling ratio $C=\chi/\sigma^2$ versus $\sigma$.
The dotted lines show the linear fits.
}
\label{chi2}
}

\section{Results}
\label{sec3}

In this section we present and discuss the Monte Carlo results
concerning the topological susceptibility $\chi$ and the function
$s(\theta)$, cf. Eq.~(\ref{sdef}).   
In Tables~\ref{tableres3}, \ref{tableres4} and \ref{tableres6} we report $\chi$,
the ratio $C=\chi/\sigma^2$, and the coefficient $b_2$, 
cf. Eq.~(\ref{b2exp}).
Data for different lattice sizes allow us to check
for finite size effects: their comparison shows that they
are under control.
The ratio $C$ is plotted in Fig.~\ref{chi2} versus $\sigma$,
to evidentiate possible scaling corrections, which
are expected to be $O(a^2)$ apart from logarithms.  
Such corrections are indeed clearly observed for $N=3$
and $N=4$, and one needs to allow for them in the fitting procedure
to extrapolate to the continuum limit.
For this purpose we assume a linear behavior 
\begin{equation}
C= C_{\rm cont} + u \sigma, 
\end{equation}
which takes into account of the leading
$O(a^2)$ scaling  correction and it is also supported by the data.
We obtain
\begin{eqnarray}
& C_{\rm cont} = 0.0282(12),\quad u=0.077(24) \qquad & {\rm for}\quad N=3, \\
& C_{\rm cont} = 0.0257(10),\quad u=0.049(15)  \qquad & {\rm for}\quad N=4. 
\end{eqnarray}
For $N=6$ no evidence of scaling corrections is observed.
Therefore, in this case  as final estimate we take the weighted average of the
data for the largest values of $\beta,\gamma$,
i.e. for $\gamma=0.348$ and lattice size $L=12,16$ and for $\gamma=0.350$, 
obtaining
\begin{equation}
C_{\rm cont} = 0.0236(10) \qquad {\rm for}\quad N=6.
\end{equation}
Results for smaller $\gamma$ are used to control
scaling corrections: they are perfectly consistent,
indicating that scaling corrections are at most of 
the size of the error
\footnote{
We mention that a fit of the $N=6$ data to a constant 
gives $C_{\rm cont}=0.0232(5)$ with $\chi^2/d.o.f.\approx 0.5$,
while a linear fit to $a+b\sigma$ gives
$C_{\rm cont}=0.0240(25)$ and $u=-0.01(3)$.}.

These results of $C_{\rm cont}$ are plotted in Fig.~\ref{summary}
versus $1/N^2$, which is the expected order of the
asymptotic corrections in the large-$N$ limit.
They are clearly consistent with a behavior of
the type 
\begin{equation}
C_{\rm cont}(N) = C_\infty + {B\over N^2}.
\label{Nm2corr}
\end{equation}
Assuming such a behavior,
a fit to the results for $N=3,4,6$ gives
\begin{equation}
C_\infty = 0.0221(14), \qquad B=0.055(18),
\label{ourresults}
\end{equation}
which provides a rather accurate estimate of the
large-$N$ limit of $C$; this limit is clearly seen to be
nonzero.
Note that, extrapolating using only the $N=4,6$ results, we
obtain a perfectly consistent result, i.e.
$C_{\rm cont}=0.0219(20)$ and $B=0.06(4)$.

In order to compare with the numbers appearing in the
literature, we translate the above results in physical units assuming
the rather standard
value $\sqrt{\sigma}=440 \, {\rm MeV}$, for $N \geq 3$. We obtain
$\chi_\infty^{1/4}=170(3) \, {\rm MeV}$, and $\chi^{1/4}=180(2) \,{\rm
MeV}$ for $N=3$.  These numbers have to be compared with the r.h.s. of
Eq.~(\ref{wittenformula}), which, using the actual values of $f_\pi$,
$m_{\eta '}$, and $N_f=3$, gives $( 191 \, {\rm MeV})^4$; using the
formula refined by Veneziano \cite{Veneziano-79} for which $m_{\eta'}^2 \rightarrow
m_{\eta'}^2 + m_\eta^2 - 2 m_K^2$, one obtains $(180 \, {\rm MeV})^4$.
Our results for $N=3$ are in substantial agreement with the estimates
obtained so far using various methods, see, e.g.,
Refs. \cite{ADD-97,ST-98,HN-98,EHN-98,GHS-02,CTW-02}. 
Ref.~\cite{LT-01} presents an estimate of $C_\infty$ obtained
by extrapolating results for $N=2,3,4,5$ using the formula (\ref{Nm2corr});
they obtain $C_\infty=0.0200(43)$ from $C_{\rm cont}(N=2)=0.0545(25)$, 
$C_{\rm cont}(N=3)=0.0355(33)$, $C_{\rm cont}(N=4) = 0.0224(39)$ 
and $C_{\rm cont}(N=5) = 0.0224(49)$.
Our work improves these results essentially for two reasons:
we performed simulations with much higher statistics,
and we do not rely on $N=2$ data for our extrapolation, which may be 
too small a value of $N$.
The large-$N$ extrapolation of Ref.{~\cite{LT-01}
is in agreement with ours, although we note some differences in the 
finite $N$ results.
The extrapolation of Ref.~\cite{LT-01} strongly depends
on the $N=2$ result, while their $N=2$ result is not fitted by our extrapolation
(\ref{ourresults}), which would give approximately 0.036 for $N=2$.
Thus, at this stage
the agreement of the large-$N$ extrapolations appears to be rather accidental.
The result for $N=2$ of Ref.~\cite{LT-01} would
suggest that $N=2$ is outside the region
where the behavior (\ref{Nm2corr}) sets in.

\FIGURE[ht]{
\epsfig{file=summary.eps, width=12truecm} 
\caption{
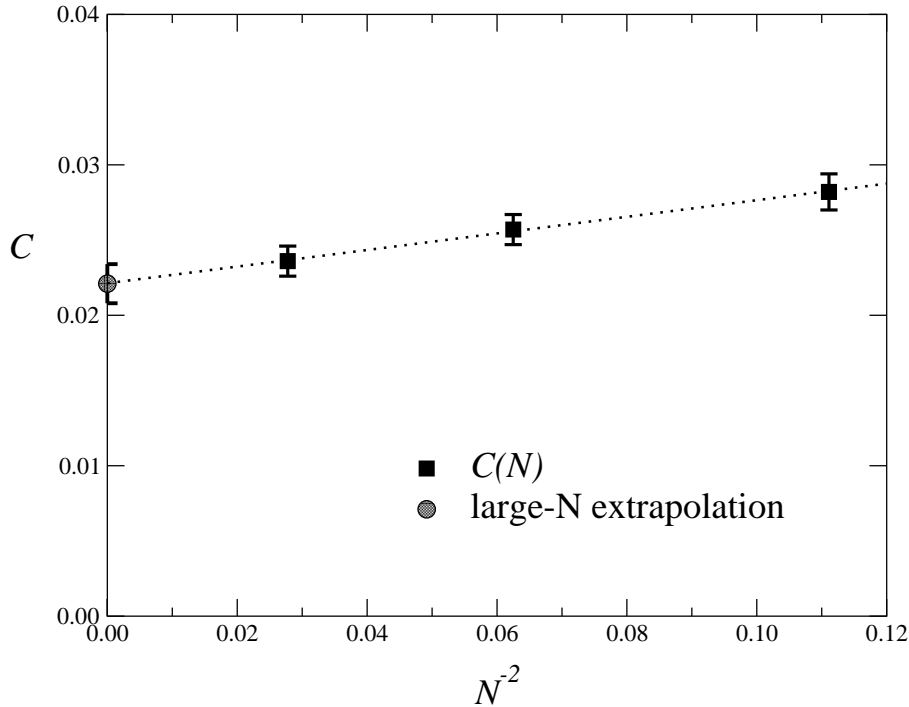
The scaling ratio $C=\chi/\sigma^2$ versus $1/N^2$.
The dotted line shows the linear fit to $A+B/N^2$.
}
\label{summary}
}

Let us now turn to the coefficient $b_2$
in the expansion (\ref{sexp}) of the function $s(\theta)$,
which determines the correction of $F(\theta)$ from 
the simple $\theta^2$ behavior.
Note that a reasonably accurate computation of $\chi_4$, which is necessary to determine
$b_2$,  requires very high
statistics due to a large cancellation between the
two terms in its definition (\ref{chi4}).
Sufficiently accurate  results have been obtained only in the
cases $N=3,4$. 
The results are listed in Tables~\ref{tableres3} and
\ref{tableres4} for $N=3,4$ respectively.
For $N=6$ they were always consistent 
with zero but with a large error, essentially due
to the much smaller statistics available, taking also into account
the critical slowing down phenomenon discussed before.
We report only the most precise result
obtained by a high-statistics simulation:
$b_2=-0.01(2)$ for $\gamma=0.348$ and $L=12$.
The results for $N=3$ and $N=4$ are plotted in Fig.~\ref{b2plot}.
They show scaling within the errors. 
Assuming that the scaling corrections are negligible with respect to
the typical errors of the data,
as final estimate we consider
the weighted average of the data, and as error we quote their typical
error (in this computation we discard the 
data for $\beta=6.0$ and $L=12$ because of possible finite size effects).
We obtain  
\footnote{
Fitting the data to a constant, 
one obtains $b_2=-0.0226(34)$ for $N=3$ 
and $b_2=-0.0125(36)$ for $N=4$, with acceptable $\chi^2$.
These errors are substantially smaller 
than those shown in Eqs.~(\ref{b2n3}) and (\ref{b2n4}).
We believe that the final errors reported in the text
are more reliable, since scaling
corrections appear to be negligible only within
the typical errors of the data.
Nevertheless, we note that the $1/N^2$ behavior is verified
even within the smaller errors of the fits.
}
\begin{eqnarray}
& b_2 = - 0.023(7),\qquad & {\rm for}\quad N=3, \label{b2n3} \\
& b_2 = - 0.013(7),\qquad & {\rm for}\quad N=4. \label{b2n4}
\end{eqnarray}
For $N=6$ we report the estimate $b_2=-0.01(2)$ obtained
at $\gamma=0.348$. 
These results show that $b_2$ is very small and
are consistent with the hypothesis $b_2=O(1/N^2)$.
One may also assume 
$b_2\approx b_{2,2}/N^2$ and fit the above results,
obtaining $b_{2,2}=-0.21(5)$.

\FIGURE[ht]{
\epsfig{file=b2.eps, width=12truecm} 
\caption{
The scaling ratio $b_2$ versus $\sigma$.
}
\label{b2plot}
}

As already mentioned, we used a cooling technique to
determine the topological charge of a lattice configuration.
Several numerical studies have shown that cooling methods provide
quite reliable results. Nevertheless,
some of the arguments on which cooling relies are not rigorous. Therefore
it would be worthwhile to confirm our results using alternative
methods.
A promising and more rigorous approach is the one based on the
index theorem for Ginsparg-Wilson fermions \cite{Luscher-98,HLN-98,GRTV-01},
using, e.g., the overlap lattice fermion formulation
\cite{Neuberger-98}.

In conclusion, we have computed the $\theta$ dependence
of the ground-state energy of four-dimensional SU($N$) gauge theories
around $\theta=0$, 
by evaluating numerically the coefficients of its 
expansion to $O(\theta^4)$.
We have verified that the coefficient
of the $\theta^2$ term, which is the topological susceptibility,
has a nonzero large-$N$ limit, in agreement
with the picture proposed by Witten \cite{Witten-79}
to provide an explanation to the U(1)$_A$ problem. 
Corrections to the leading $\theta^2$ term turn
out to be very small. Indeed,
the coefficient $b_2$ of the $\theta^4$ term is
a few per cent for $N=3$, and appears to decrease 
with increasing $N$, consistently with
Witten's conjecture \cite{Witten-98} implying
$b_2=O(1/N^2)$. 
Clearly, higher-order terms in the expansion in powers of $\theta^2$
might not be suppressed by powers of $N$ and therefore might spoil the
simple $\theta^2$ behaviour predicted by Witten for large $N$. The study of
these terms requires the measurement of higher moments of the
topological charge probability distribution and are beyond the scope
of this work.  Thus, for $N\geq 3$ the simple Gaussian form
\begin{equation}
f(\theta)\approx {1\over 2} C \theta^2
\end{equation}
is expected to provide a good approximation of the dependence on
$\theta$ of the scaling free energy, up to $O(\theta^6)$ terms,
and therefore for a relatively large range of values of $\theta$.

\acknowledgments{ We thank 
Maurizio Davini for his valuable and indispensable technical support.
H. P. would like to thank the Theory Group in Pisa for their
hospitality during various stages of this work. We thank G. Veneziano
for helpful comments.}

\end{document}